\documentclass[10pt,twocolumn,showpacs,longbibliography,prb,aps,floatfix]{revtex4-1}

\usepackage{graphicx}
\usepackage{amsmath}
\usepackage{amssymb}
\usepackage{bm}
\usepackage{relsize}

\newcommand{\Eg}{\bar{E}_g}
\newcommand{\beps}{\bar{\varepsilon}}
\newcommand{\bom}{\bar{\omega}}
\newcommand{\kz}{\bar{k}_z}
\newcommand{\nf}{\bar{n}_f}
\newcommand{\Nf}{\bar{N}_{f,\xi}}

\renewcommand{\Re}{\text{Re }}
\renewcommand{\Im}{\text{Im }}

\newcommand{\om}{\omega}
\newcommand{\eps}{\varepsilon}
\newcommand{\Del}{\Delta}
\newcommand{\del}{\delta}
\newcommand{\Th}{\Theta}
\newcommand{\kap}{\kappa}

\begin{document}

\title{An analytic evaluation of Kane fermion magneto-optics in two and three dimensions}

\author{J.D. Malcolm}
\author{E.J. Nicol}
\affiliation{Department of Physics, University of Guelph, Guelph, Ontario N1G 2W1 Canada} 
\affiliation{Guelph-Waterloo Physics Institute, University of Guelph, Guelph, Ontario N1G 2W1 Canada}
\date{\today}

\begin{abstract}
{We calculate and present an analytic form of the magneto-optical conductivity for the gapped low-energy Kane model in two and three dimensions separately.  The two-dimensional case maps onto the $\alpha$-$\mathcal{T}_3$ model at a particular value of $\alpha=1/\sqrt{3}$.  In two dimensions, two chiral sectors exist, between which there are no optically activated transitions.  In three dimensions, the extra dimension of dispersion mixes the two sectors so that intra- and inter-sector transitions can occur.  The latter type of transition can be separated out via circular polarizations of light and shows a distinct signature in the transverse conductivity.}
\end{abstract}


\maketitle

\section{Introduction}\label{sec:Intro}

\begin{figure}
\begin{center}
\includegraphics[width=1.0\linewidth]{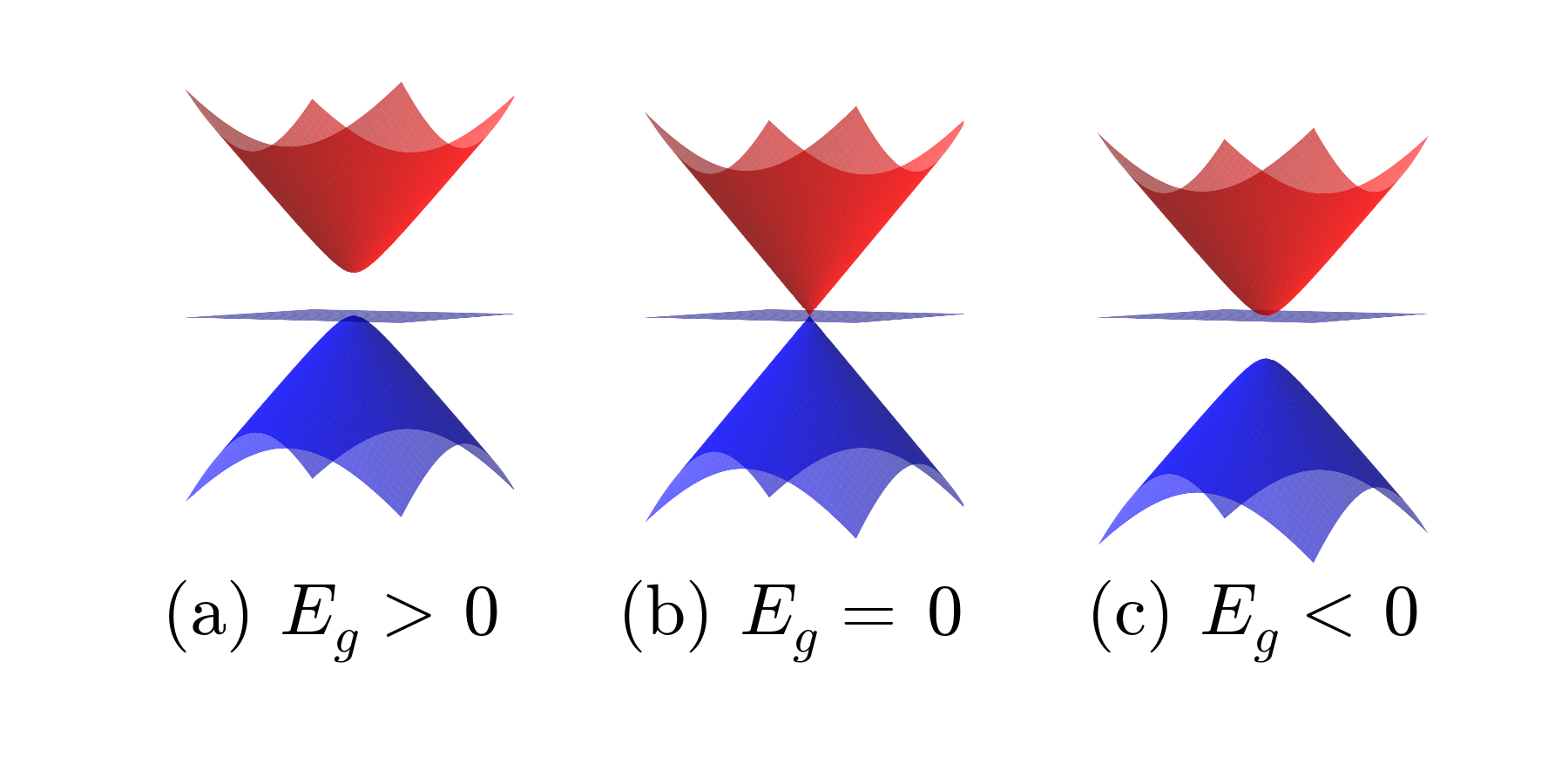}
\end{center}
\caption{\label{fig:GapCones}(Color online)  Band structure of the low-energy Kane model showing its evolution through tuning of the gap $E_g$.  A chemical potential $\mu=0^+$ fills the lower cone and the flat band (blue), leaving the upper cones unoccupied (red).}
\end{figure}

The Kane model,\cite{Kane56} used to calculate the band structure of zinc-blende semiconductors, has recently gained interest in the broad study of relativistic materials, which concerns systems with single-point linear band crossings.  The term relativistic here does not refer to ultra-fast speeds in the material, but rather that quasiparticle behaviour can be described by equations of motion typical of relativistic particles (massless Dirac fermions, etc.).  At low energy, the three-dimensional (3D) model consists of two conic bands and a nominally-flat heavy-hole band (Fig.~\ref{fig:GapCones}).  A tunable gap parameter, $E_g$, takes the system from semiconductor ($E_g>0$, Fig.~\ref{fig:GapCones}(a)) to semimetal ($E_g<0$, Fig.~\ref{fig:GapCones}(c)).  At the critical value $E_g=0$, the upper and lower cones linearize and touch at a single point intersecting the flat band (Fig.~\ref{fig:GapCones}(b)), showing the same dispersion as the pseudospin-1 Weyl semimetal.\cite{Burkov11,Delplace12}

The Kane model (detailed in Sec.~\ref{sec:KaneModel}) is particularly interesting on two fronts.  The first is mathematical.  We have shown previously\cite{Malcolm15} that when restricted to two dimensions (2D), the Kane model maps onto an intermediate value ($\alpha=1/\sqrt{3}$) of the $\alpha$-$\mathcal{T}_3$ model\cite{Raoux14} which interpolates between the pseudospin-1/2 (graphene, $\alpha=0$) and pseudospin-1 (dice or $\mathcal{T}_3$ lattice, $\alpha=1$) Dirac-Weyl systems.  This makes the 2D Kane fermion a hybrid of a pseudospin-1 and -1/2 Dirac fermion.  The $\alpha$-$\mathcal{T}_3$ has itself gathered much attention recently with theorists as an illuminating toy model.\cite{Dora14,Piechon15,Louvet15,Illes15,Illes16,Biswas16}

On the physical front, optical measurements have shown that bulk Hg$_{1-x}$Cd$_x$Te (MCT) at low temperature and a critical doping of $x=x_c\approx0.17$ exhibits so-called massless Kane fermions.\cite{Orlita14}  That is, this particular phase is described by the gapless Kane model.  There is also evidence that, in addition to doping, the gap in MCT can be tuned open and shut with temperature.\cite{Teppe16}  More recent optics observations by the same group also reveal the presence of massless Kane fermions in Cd$_3$As$_2$.\cite{Akrap16}

The magneto-optical conductivity of a system is an important and useful tool in identifying the quasiparticle behaviour in a material.  Signatures of relativistic fermions are displayed in the locations and heights of peaks in the optical spectrum, and in the peaks' modification through changes in magnetic field strength, $B$, and chemical potential, $\mu$.\cite{Malcolm14}  With elevated attention focused on massless Kane fermions, and with magneto-optical measurements being the main tool used in their identification, there is a strong motivation for theoretical investigations into the Kane magneto-optics.

In a previous paper, we used an extended Kane model (which includes a fourth band at large negative energies not shown in Fig.~\ref{fig:GapCones}) to calculate the absorbance of MCT in the presence of a large magnetic field in order to match a specific set of experimental data (obtained under a 16 T field).\cite{Orlita14,Malcolm15}  The large magnetic field made it necessary to use the extended model, which could only be solved numerically.

In this current work, we focus on the reduced Kane model (appropriate for lower fields) which allows for the derivation of analytic expressions for the Kane magneto-optics with dependence on a tunable gap, $E_g$, the chemical potential, $\mu$, and temperature, $T$.  Analytic derivation allows for deep insights into the origin of certain features seen in optical spectra and the resulting mathematical expression gives the conductivity's explicit dependence on various parameters.  In addition, these analytic formulas are now readily available to experimentalists working on relevant systems at low magnetic field.   The following analysis is done for both the Kane model restricted to 2D (Sec.~\ref{sec:2D}) and open to the full 3D (Sec.~\ref{sec:3D}).  Investigation of the 2D system is relevant to research surrounding other 2D relativistic systems\cite{Raoux14,Dora14} and also provides a useful reference when describing aspects of the 3D system.

\section{Kane Model}\label{sec:KaneModel}

The low-energy Kane model consists of two 2D sectors, A and B, which are linked by a dimension of dispersive momentum, $\hbar k_z$.  The Hamiltonian is
\begin{equation}\label{eqn:Ham3D}
\hat{\mathcal{H}} = 
\begin{pmatrix}
\hat{\mathcal{H}}_A & \hbar vk_z\hat{C} \\
\hbar vk_z\hat{C}^T & \hat{\mathcal{H}}_B \\
\end{pmatrix}\,,
\end{equation}
where, in the presence of a magnetic field $\bm{B}=B\hat{e}_z$,
\begin{align}
& \hat{\mathcal{H}}_A = \gamma
\begin{pmatrix}
0 & \sqrt{3}a & 0 \\
\sqrt{3}a^\dagger & \Eg & -a \\
0 & -a^\dagger & 0 \\
\end{pmatrix}\,,\label{eqn:HamA} \\
& \hat{\mathcal{H}}_B = \gamma
\begin{pmatrix}
0 & a & 0 \\
a^\dagger & \Eg & -\sqrt{3}a \\
0 & -\sqrt{3}a^\dagger & 0 \\
\end{pmatrix}\,,\label{eqn:HamB} \\
& \hat{C} = 
\begin{pmatrix}
0 & 0 & 0 \\
-1 & 0 & 0 \\
0 & -1 & 0 \\
\end{pmatrix}\,.
\end{align}
$\gamma=\hbar v/\sqrt{2}\ell_B$ is an energy factor with $\ell_B = \sqrt{\hbar/eB}$ the magnetic length scale.  $\bar{x}=x/\gamma$ wherever this notation occurs, $v$ is a parameter characteristic of the material with dimensions of velocity, and $[a,a^\dagger]=1$ are the Fock-space operators associated with the quantized Landau levels (LL) formed in a finite magnetic field.  When restricted to 2D by forcing $k_z=0$, the Hamiltonian reduces into two independent sectors,
\begin{equation}\label{eqn:Ham2D}
\hat{\mathcal{H}}_{2D} = \hat{\mathcal{H}}_A\oplus\hat{\mathcal{H}}_B\,.
\end{equation}

\begin{figure}
\begin{center}
\includegraphics[width=1.0\linewidth]{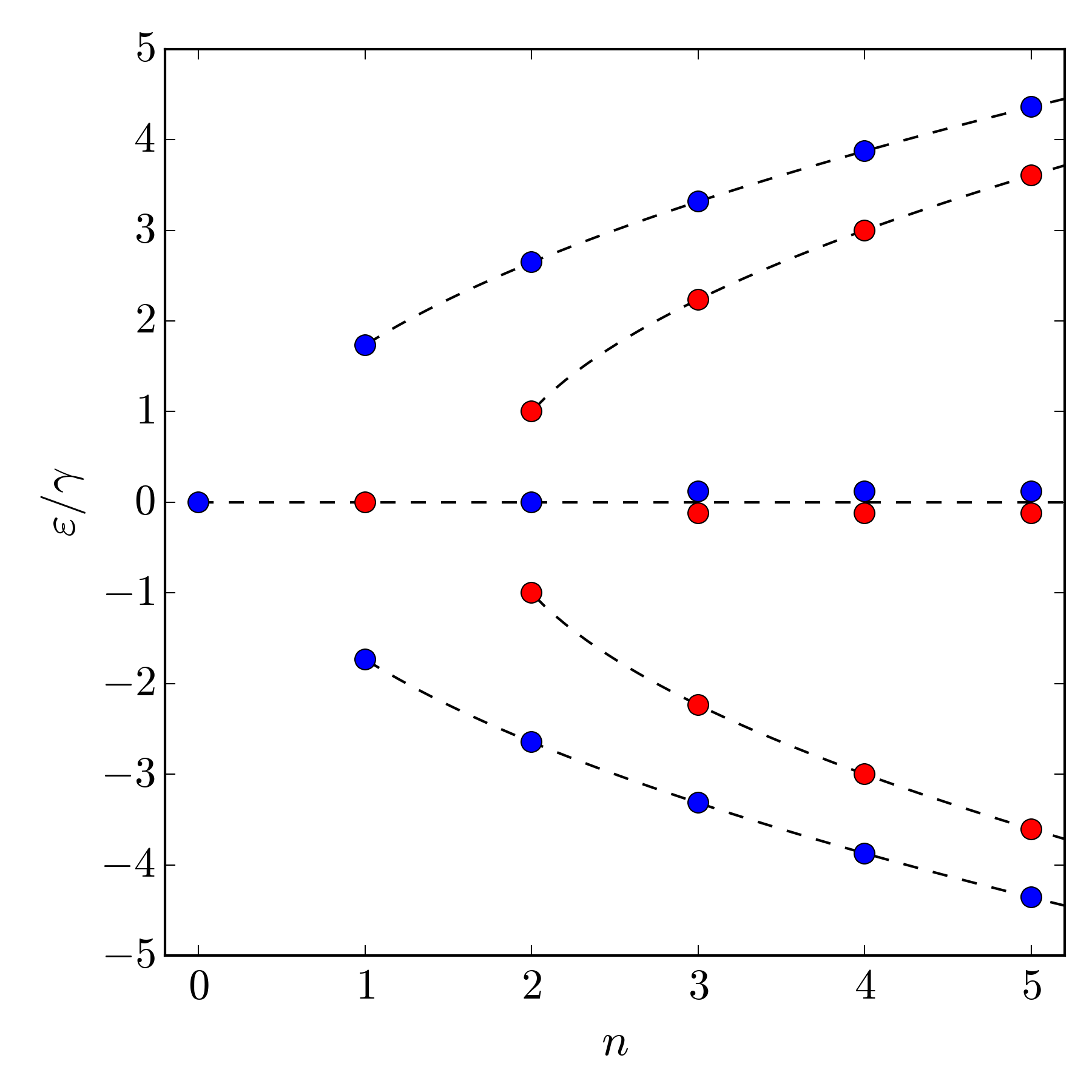}
\end{center}
\caption{\label{fig:BareSnowshoe}(Color online) Landau level energy dependence on Fock number $n$ of the Kane model at $k_z=0$ with zero gap, $E_g=0$.  Red levels reside in sector A and blue in sector B.  Dashed lines trace out the branches off of which the Landau levels bud.  The $n=3,4,5$ levels along the flat branch are only offset here for illustration.}
\end{figure}

As mentioned in the previous section, the 2D Kane system in Eq.~(\ref{eqn:Ham2D}) is exactly the $\alpha$-$\mathcal{T}_3$ model with $\alpha=1/\sqrt{3}$.  In that model the value $\alpha=0$ ($\alpha=1$) describes the physics of graphene (dice lattice), where $H_A$ and $H_B$ pertain to the distinct chiral centres K and K', respectively.\cite{Raoux14,Dora14,Illes15,Illes16,Kovacs16}  In discussions of 3D Kane fermions,\cite{Orlita14,Teppe16,Akrap16} these two sectors are instead typically given the label $\sigma=\pm$ and are referred to as a Kramer's spin pair.  In the 3D case however, non-zero $k_z$ mixes the two sectors via the connection $\hat{C}$.  Thus, any such label on a state only truly describes a distinct chirality, or Kramer's spin, at $k_z=0$.  This property is somewhat analogous to Dirac fermions in 3D, where it is a finite-valued mass that takes the role of connecting two otherwise independent chiral sectors.\cite{Peskin95}

Equation~(\ref{eqn:Ham3D}) is a low-energy approximation of the Kane model.  For large magnetic fields, where LL's are pushed up to higher energies, this Hamiltonian is insufficient to represent MCT.\cite{Orlita14,Malcolm15}  When modelling frequency-dependent the magneto-optical spectrum of MCT under a 16 T magnetic field in a previous publication, it was necessary to expand Eq.~(\ref{eqn:Ham3D}) from a $6\times6$ matrix to the proper $8\times8$ matrix.\cite{Orlita14,Malcolm15}  This expansion considers the effect of a fourth band, which is well separated from the others.  While this band does not introduce charge carriers at the energies studied, it induces curvature in the conic bands, altering their energies and wavefunctions.  The magneto-optical conductivity for the $8\times8$ system must be solved numerically.  With the low-energy Hamiltonian in Eq.~(\ref{eqn:Ham3D}) however, the conductivity can be solved analytically, which is the focus of this paper.

\section{Magneto-optics in Two Dimensions}\label{sec:2D}

In this section, we provide the magneto-optical conductivity for the Kane model restricted to two dimensions (Eq.~(\ref{eqn:Ham2D})).  This has already been done via the general $\alpha$-$\mathcal{T}_3$ model for zero gap.\cite{Illes16}  We have also recently become aware of the same calulation for the gapped $\alpha$-$\mathcal{T}_3$ model by a method using the spectral representation.\cite{Kovacs16}  Our presentation of the 2D magneto-optics here is for completeness and to supplement the main focus of our work, which is the 3D magneto-optics.

\begin{figure}
\begin{center}
\includegraphics[width=1.0\linewidth]{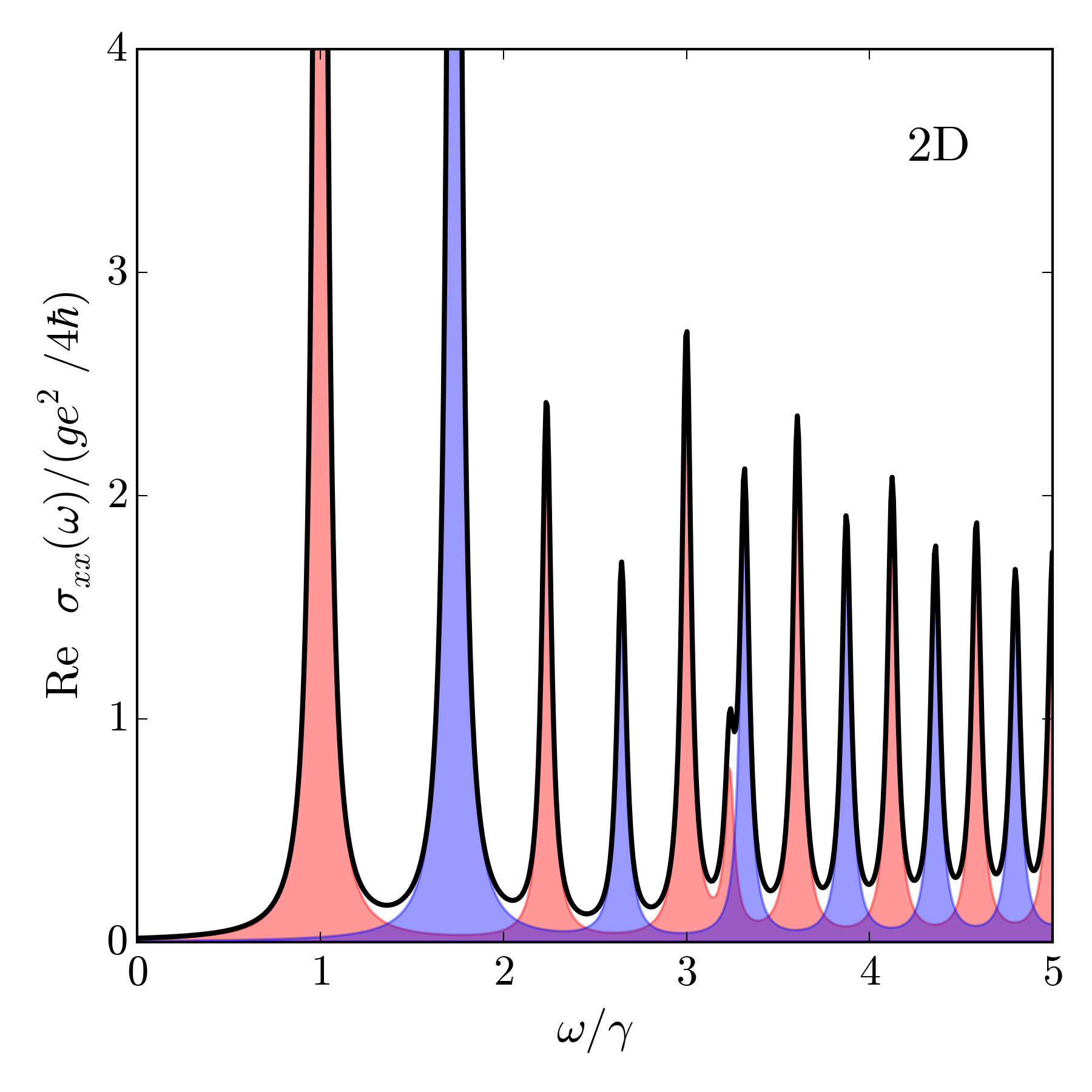}
\end{center}
\caption{\label{fig:Cond2D}(Color online) Longitudinal part of the magneto-optical conductivity in the 2D Kane model (black).  Contributions from sector A are shown in red and from sector B in blue.  Parameters are $T=0$, $E_g=0$, $\mu=\gamma/2$, and $\eta=0.03\gamma$.}
\end{figure}

\begin{figure}
\begin{center}
\includegraphics[width=1.0\linewidth]{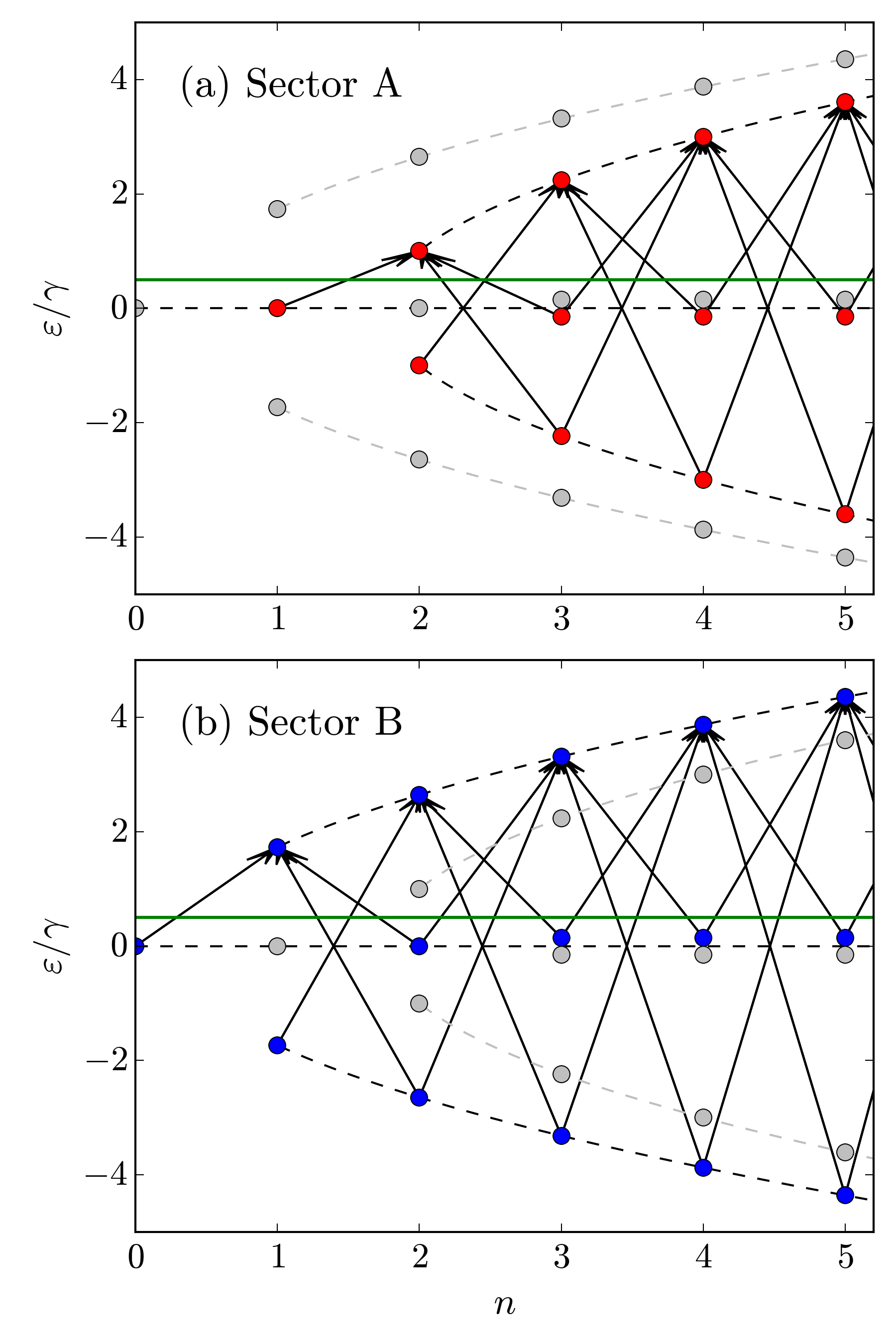}
\end{center}
\caption{\label{fig:Snowshoe2D}(Color online) Snowshoe diagrams for (a) sector A and (b) sector B showing the allowed transitions between Landau levels for the chemical potential $\mu=\gamma/2$ shown in green.  The spectrum in each panel is the same as in Fig.~\ref{fig:BareSnowshoe} with the uninvolved sector shaded to gray.}
\end{figure}

In 2D, sectors A and B are independent, meaning their contributions to the conductivity can be considered separately.  In the presence of a magnetic field, the continuous dispersion in Fig.~\ref{fig:GapCones} condenses into discrete LL's (Fig.~\ref{fig:BareSnowshoe}) labelled by Fock number $n$.  The eigensystem for sector A, giving LL energies and their associated wavefunctions, is
\begin{equation}\label{eqn:eigenA}
\begin{split}
& \beps^A_{0,n} = 0 \;,\; n\geq1\;(n\neq2)\,; \\
& \beps^A_{\xi,n} = \big(\Eg/2\big)+\xi\sqrt{\big(\Eg/2\big)^2+4n-7} \;,\; n\geq2\,; \\
& \psi^A_{0,n} = \kap\Big(\sqrt{n-1}|n-3\rangle,0,\sqrt{3(n-2)}|n-1\rangle\Big)^T\,; \\
& \psi^A_{\xi,n} = \kap\Big(\sqrt{3(n-2)}|n-3\rangle,\beps^A_{\xi,n}|n-2\rangle, \\
& \phantom{xxxxxxxxxxxxxxxxxxxxx} -\sqrt{n-1}|n-1\rangle\Big)^T\,;
\end{split}
\end{equation}
and the same for sector B is
\begin{equation}\label{eqn:eigenB}
\begin{split}
& \beps^B_{0,n} = 0 \;,\; n\geq0\;(n\neq1)\,; \\
& \beps^B_{\xi,n} = \big(\Eg/2\big)+\xi\sqrt{\big(\Eg/2\big)^2+4n-1} \;,\; n\geq1\,; \\
& \psi^B_{0,n} = \kap\Big(\sqrt{3n}|n-2\rangle,0,\sqrt{n-1}|n\rangle\Big)^T\,; \\
& \psi^B_{\xi,n} = \kap\Big(\sqrt{n-1}|n-2\rangle,\beps^B_{\xi,n}|n-1\rangle,-\sqrt{3n}|n\rangle\Big)^T\,;
\end{split}
\end{equation}
where $|m\!\!<\!\!0\rangle=0$, $a|m\rangle=\sqrt{m}|m-1\rangle$, $a^\dagger|m\rangle=\sqrt{m+1}|m+1\rangle$, and $\kap$ is a normalization factor (different for each vector).  The subscript indices 0 and $\xi=\pm$ are reminiscent of the band labels in the zero-field system for the flat band, and upper- and lower-cones, respectively.  With the discrete quantization of the continuous bands under a magnetic field, it is more appropriate to refer to these as labels for the branches off of which the LL's bud.  The LL energies over Fock number $n$ are plotted in Fig.~\ref{fig:BareSnowshoe} with sector A in red and sector B in blue.  It is admittedly cumbersome to have the sector A Fock index begin at $n=1$ instead of $n=0$, like in sector B.  However,  the three-element wavefunctions in Eqs.~(\ref{eqn:eigenA}) and~(\ref{eqn:eigenB}) are decomposed from the six-element 3D wavefuntion at $k_z=0$ (Sec.~\ref{sec:3D}).  Thus, in order to remain consistent with the 3D system, we keep this indexing convention in the wavefunctions.

The magneto-optical conductivity tensor is calculated via the Kubo formula in the LL representation.\cite{Mahan13}  In 2D this is
\begin{equation}\label{eqn:Kubo2D}
\sigma^{\rm 2D}_{\alpha\beta}(\om) = \frac{ig}{2\pi\hbar\ell_B^2}\sum_{\psi,\psi'}\frac{\Del n_f}{\Del\eps}\frac{\langle\psi|\hat{\jmath}_\alpha|\psi'\rangle\langle\psi'|\hat{\jmath}_\beta|\psi\rangle}{\om-\Del\eps+i\eta}\,,
\end{equation}
where $\alpha,\beta=\{x,y\}$, $g$ is a degeneracy factor ($g=2$ in the Kane model for the two-fold intrinsic spin degeneracy), $\hat{\jmath}_\alpha$ are the current operators, $\eta=0^+$ can be considered as an infinitesimal scattering rate, and $\Del\eps=\eps'-\eps$.  The factor $\Del n_f=n_f(\eps)-n_f(\eps')$ ensures that only scattering events from a filled state $\psi$ (with energy $\eps$) to an unoccupied state $\psi'$ are included, where $n_f(x)$ is the Fermi-Dirac distribution at temperature $T$ and chemical potential $\mu$.  Because Eq.~(\ref{eqn:Ham2D}) is reducible to two independent sectors, Eq.~(\ref{eqn:Kubo2D}) can be calculated for each sector A and B separately and then summed.  Using the Kane model current operators $\hat{\jmath}_\alpha=e\cdot\partial\hat{\mathcal{H}}/\partial k_\alpha$ obtained from Eqs.~(\ref{eqn:HamA}) and~(\ref{eqn:HamB}), the absorptive part of the conductivity is
\begin{equation}\label{eqn:cond2DKane}
\begin{split}
& \left.
\begin{aligned}
\Re \sigma^{\rm 2D}_{xx}(\om) \\
\Im \sigma^{\rm 2D}_{xy}(\om)
\end{aligned}
\right\}
= \frac{ge^2}{4\hbar}\sum_{\psi,\psi'}\frac{\Del\nf}{\Del\beps}\del(\bom-\Del\beps) \\
&\phantom{xxxxxxx} \times\bigg[\del_{n',n-1}\big|f(\psi,\psi')\big|^2\pm\del_{n',n+1}\big|f(\psi',\psi)\big|^2\bigg]\,,
\end{split}
\end{equation}
where $\nf(\bar{x})=n_f(x)$.  The overlap functions in the respective sectors are
\begin{align}
\label{eqn:OverlapA}& f^A(\psi,\psi') = \sqrt{3}\alpha_1\alpha_2'-\alpha_2\alpha_3'\,, \\
\label{eqn:OverlapB}& f^B(\psi,\psi') = \alpha_1\alpha_2'-\sqrt{3}\alpha_2\alpha_3'\,,
\end{align}
where $\alpha_i$ is the $i^{\rm th}$ element of vector $\psi$.  Equation~(\ref{eqn:cond2DKane}) reflects the strict $n\rightarrow n\pm1$ dipole selection rule in optically activated transitions between LL's.  Contributions to the conductivity can be broken down further into flat-cone (FC) and cone-cone (CC) transitions, so that
\begin{equation}\label{eqn:Cond2DSum}
\left.
\begin{aligned}
\Re \sigma^{\rm 2D}_{xx}(\om) \\
\Im \sigma^{\rm 2D}_{xy}(\om)
\end{aligned}
\right\}
= \frac{ge^2}{4\hbar}\sum_{\zeta=A,B}\bigg[\big(FC\big)^\zeta_{\substack{\\xx\\xy}} + \big(CC\big)^\zeta_{\substack{\\xx\\xy}}\bigg]\,.
\end{equation}
Note that CC includes both intra- and intercone transitions.

The results for the 2D Kane model magneto-optical conductivity are
\begin{widetext}
\begin{equation}\label{eqn:2D_FCA}
\big(FC\big)^A_{\substack{\\xx\\xy}} = \frac{3}{2}\sum_{\xi=\pm}\sum_{n=0}^\infty\big(\del_{\xi,+}\pm\del_{\xi,-}\big)\big|\nf(0)-\nf(\beps^A_{\xi,n+2})\big|\bigg(\frac{n+2}{4n+5}\pm\frac{n-1}{4n-3}\bigg)\frac{\del\big(\bom-|\beps^A_{\xi,n+2}|\big)}{\sqrt{\big(\Eg/2\big)^2+4n+1}}\,,
\end{equation}
\begin{equation}
\begin{split}
\big(CC\big)^A_{\substack{\\xx\\xy}} = \frac{1}{4}\sum_{\xi,\xi'=\pm}\sum_{n=0}^\infty\big(\del_{\xi',-}\pm\del_{\xi',+}\big)\big|\nf(\beps^A_{\xi,n+2})-\nf(\beps^A_{\xi',n+3})\big|\frac{\del\big(\bom-|\beps^A_{\xi',n+3}-\beps^A_{\xi,n+2}|\big)}{|\beps^A_{\xi',n+3}-\beps^A_{\xi,n+2}|} \phantom{xxxxxx} \\
\times\frac{(n+1)\big|6\xi\xi'+|\beps^A_{\xi,n+2}|^{-1}+9|\beps^A_{\xi',n+3}|^{-1}\big|}{\sqrt{\big[\big(\Eg/2\big)^2+4n+1\big]\big[\big(\Eg/2\big)^2+4n+5\big]}}\,,
\end{split}
\end{equation}
\begin{equation}
\big(FC\big)^B_{\substack{\\xx\\xy}} = \frac{3}{2}\sum_{\xi=\pm}\sum_{n=0}^\infty\big(\del_{\xi,+}\pm\del_{\xi,-}\big)\big|\nf(0)-\nf(\beps^B_{\xi,n+1})\big|\bigg(\frac{n+2}{4n+7}\pm\frac{n-1}{4n-1}\bigg)\frac{\del\big(\bom-|\beps^B_{\xi,n+1}|\big)}{\sqrt{\big(\Eg/2\big)^2+4n+3}}\,,
\end{equation}
\begin{equation}\label{eqn:2D_CCB}
\begin{split}
\big(CC\big)^B_{\substack{\\xx\\xy}} = \frac{1}{4}\sum_{\xi,\xi'=\pm}\sum_{n=0}^\infty\big(\del_{\xi',-}\pm\del_{\xi',+}\big)\big|\nf(\beps^B_{\xi,n+1})-\nf(\beps^B_{\xi',n+2})\big|\frac{\del\big(\bom-|\beps^B_{\xi',n+2}-\beps^B_{\xi,n+1}|\big)}{|\beps^B_{\xi',n+2}-\beps^B_{\xi,n+1}|} \phantom{xxxx} \\
\times\frac{(n+1)\big|6\xi\xi'+9|\beps^B_{\xi,n+1}|^{-1}+|\beps^B_{\xi',n+2}|^{-1}\big|}{\sqrt{\big[\big(\Eg/2\big)^2+4n+3\big]\big[\big(\Eg/2\big)^2+4n+7\big]}}\,.
\end{split}
\end{equation}
\end{widetext}
Each of the summations in Eqs.~(\ref{eqn:2D_FCA})-(\ref{eqn:2D_CCB}) have been shifted appropriately to begin at $n=0$.

The longitudinal part ($\Re\sigma_{xx}$) of Eq.~(\ref{eqn:Cond2DSum}) is plotted in Fig.~\ref{fig:Cond2D} for zero temperature ($T=0$), zero gap ($E_g=0$), and for a chemical potential such that the zero-energy LL's are completely filled and all positive-energy levels are unoccupied ($\mu\in(0,\gamma)$).  The contribution from sector A is shaded red and sector B in blue, while the total conductivity is given by the solid black curve.  In the figure's construction, Dirac-delta functions were substituted with Lorentzians,
\begin{equation}
\del(x)\rightarrow\mathcal{L}(x)=\frac{\eta/\pi}{x^2+\eta^2}\,,
\end{equation}
using a width $\eta=0.03\gamma$.  The conductivity in each sector shows the same pattern as the pseudospin-1 Dirac-Weyl system where peak heights from flat-to-cone transitions (the dominant contribution) decline monotonically except for the reduced-height of the second peak.\cite{Malcolm14}

\begin{figure}
\begin{center}
\includegraphics[width=1.0\linewidth]{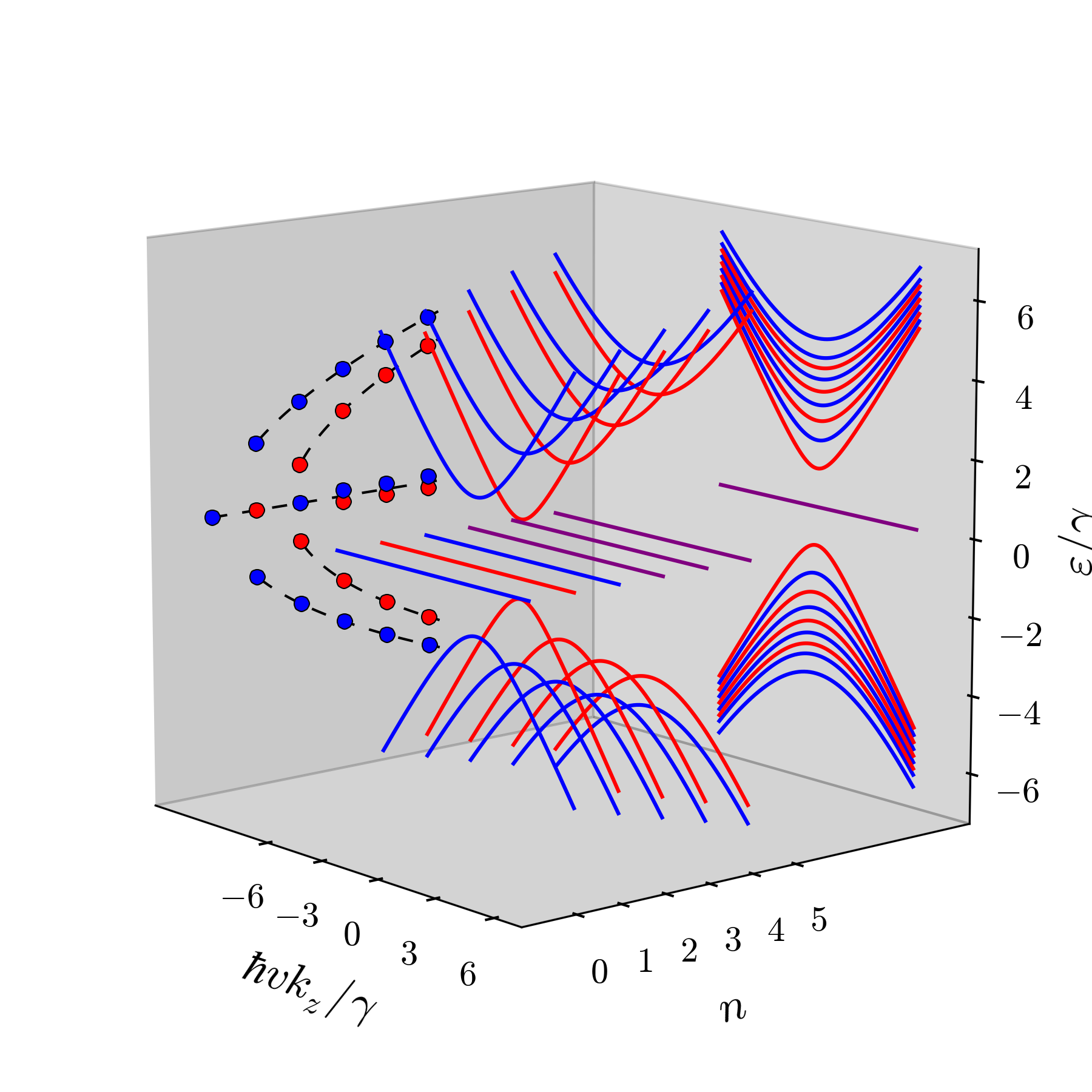}
\end{center}
\caption{\label{fig:SnowshoeProjection}(Color online) 3D Landau level energy dispersion with $n$ and $k_z$ dependence up to $n=5$.  Right-back panel shows projection along the $n$-axis, left-back panel projects out the $k_z$-extrema of each level.  Sector A levels are in red, B in blue, and overlapping zero-energy levels from different sectors are in purple.}
\end{figure}

The conductivity in Fig.~\ref{fig:Cond2D} can also be illustrated using snowshoe diagrams.\cite{Malcolm14}  A snowshoe diagram shows all of the allowed transitions between LL's for a particular $\mu$.  The sector A and sector B snowshoe diagrams are given in Figs.~\ref{fig:Snowshoe2D}(a) and~(b), respectively, with $\mu=\gamma/2$ in green marking the divide between filled and empty levels.  These diagrams are useful in providing a description of relative peak height, location, and in detailing the patterns observed with changes in chemical potential.  For example, the reduced height of the second peak in each sector (Fig.~\ref{fig:Cond2D}) comes from the fact that the second positive LL ($n=3$ for A and $n=2$ for B) is the terminus of only one transtition from the flat branch, owing to the lack of a zero-energy LL at the $n-1$ site.  All other positive LL's are the final state of two transitions from the flat branch.  In the transverse conductivity ($\Im\sigma_{xy}$), arrows with mirror symmetry about $\eps=0$ (such as the pairs of intercone transitions) make equal and opposite contributions, effectively cancelling each other out when both are present (discussed more in the following section).  The full application of snowshoe diagrams is detailed in Ref.~\cite{Malcolm14}.

\section{Magneto-Optics in Three Dimensions}\label{sec:3D}

In 3D, LL's are no longer discrete points in the Hilbert space, but disperse along $k_z$.  In addition, the two sectors A and B mix so that each LL wavefunction is a six-element vector.  The first three elements are associated with sector A and the latter three with sector B.  The LL's continue to be labelled as A or B, but the true distinction between sectors lies only at the $k_z=0$ limit.  In the 3D regime, the LL energies are
\begin{equation}\label{eqn:LLenergy}
\begin{split}
& \beps^A_{0,n}(\kz) = 0 \;,\; n\geq1\;(n\neq2)\,; \\
& \beps^A_{\xi,n}(\kz) = \big(\Eg/2\big)+\xi\sqrt{\big(\Eg/2\big)^2+\kz^2+4n-7} \;,\; n\geq2\,; \\
& \beps^B_{0,n}(\kz) = 0 \;,\; n\geq0\;(n\neq1)\,; \\
& \beps^B_{\xi,n}(\kz) = \big(\Eg/2\big)+\xi\sqrt{\big(\Eg/2\big)^2+\kz^2+4n-1} \;,\; n\geq1\,.
\end{split}
\end{equation}
The associated wavefunctions are
\begin{equation}\label{eqn:LLwavefunction}
\begin{split}
& \psi^A_{0,1} = \kap\Big(0,0,-\sqrt{3}|0\rangle,0,0,\kz|1\rangle\Big)^T\,, \\
& \psi^A_{0,n\geq3} = \kap\Big((\kz^2+n-1)|n-3\rangle,0, \\
& \phantom{x..} \sqrt{3(n-1)(n-2)}|n-1\rangle,\sqrt{3(n-2)}\,\kz|n-2\rangle,0,0\Big)^T\,, \\
& \psi^A_{\xi,n} = \kap\Big(\sqrt{3(n-2)}|n-3\rangle,\beps^A_{\xi,n}|n-2\rangle,-\sqrt{n-1}|n-1\rangle, \\
& \phantom{.....xxxxxxxxxxxxxxxxxxxxxxxxx} -\kz|n-2\rangle,0,0\Big)^T\,, \\
& \psi^B_{0,n} = \kap\Big(0,0,-\sqrt{3n}\,\kz|n-1\rangle,\sqrt{3n(n-1)}|n-2\rangle,0, \\
& \phantom{xxxxxxxxxxxxxxxxxxxxxxxxxxxx} (\kz^2+n-1)|n\rangle\Big)^T\,, \\
& \psi^B_{\xi,n} = \kap\Big(0,0,-\kz|n-1\rangle,\sqrt{n-1}|n-2\rangle,\beps^B_{\xi,n}|n-1\rangle, \\
& \phantom{xxxxxxxxxxxxxxxxxxxxxxxxxxxxxxxx}-\sqrt{3n}|n\rangle\Big)^T\,.
\end{split}
\end{equation}
For brevity in Eqs.~(\ref{eqn:LLenergy}) and~(\ref{eqn:LLwavefunction}) only, we have taken $\hbar=v=1$.  By setting $k_z=0$ in this 3D eigensystem, one recovers the same 2D eigensystem given in Eqs.~(\ref{eqn:eigenA}) and~(\ref{eqn:eigenB}).  The energy dispersion in Eq.~(\ref{eqn:LLenergy}) is plotted in Fig.~\ref{fig:SnowshoeProjection} showing $n$ and $k_z$ dependence together, up to $n=5$.  On the back-right panel, this dispersion is projected out along the $n$-axis showing only the $k_z$ dependence of the levels.  On the back-left panel, the extrema of each LL (which is located at $k_z=0$) is projected out, giving exactly the same diagram as in Fig.~\ref{fig:BareSnowshoe}.  Sector-A LL's are colored red, sector B in blue, and the overlapping zero-energy levels in purple.

\begin{figure}
\begin{center}
\includegraphics[width=1.0\linewidth]{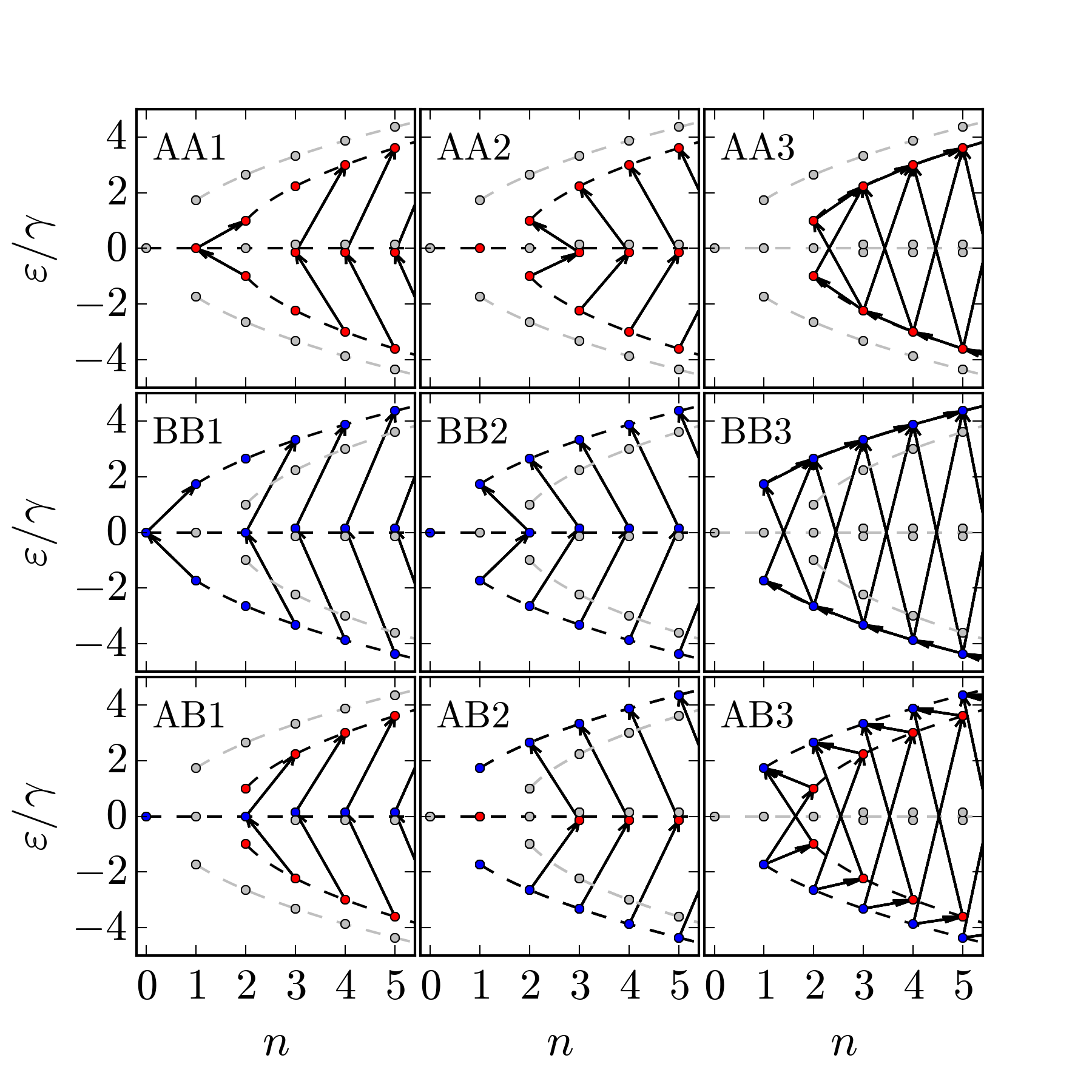}
\end{center}
\caption{\label{fig:SnowshoeParts}(Color online)  Snowshoe diagrams illustrating the complete set of transitions in each of the nine categories defined in the text.  In each panel, sector A LL's are colored red, sector B in blue, and any branches not involved in the set are shaded out to gray.}
\end{figure}

In 3D, the Kubo formula is
\begin{equation}\label{eqn:Kubo3D}
\sigma^{\rm 3D}_{\alpha\beta}(\omega) = \frac{ig}{2\pi\hbar\ell_B^2}\int_{-\infty}^\infty\frac{dk_z}{2\pi}\sum_{\psi,\psi'}\frac{\Del n_f}{\Del\eps}\frac{\langle\psi|\hat{\jmath}_\alpha|\psi'\rangle\langle\psi'|\hat{\jmath}_\beta|\psi\rangle}{\om-\Del\eps+i\eta}\,.
\end{equation}
This differs from Eq.~(\ref{eqn:Kubo2D}) by the integration over momentum.  Using the Kane model current operators,
\begin{equation}\label{eqn:Cond3DKane}
\begin{split}
& \left.
\begin{aligned}
\Re \sigma^{\rm 3D}_{xx}(\om) \\
\Im \sigma^{\rm 3D}_{xy}(\om)
\end{aligned}
\right\} = \frac{ge^2}{8\pi\hbar^2v}\frac{\gamma}{\bom}\int_{-\infty}^\infty d\kz\sum_{\psi,\psi'}\frac{\Del\nf}{\Del\beps}\del(\bom-\Del\bom) \\
& \phantom{xxxxxxxxxx} \times\bigg[\del_{n',n-1}|f(\psi,\psi')|^2\pm\del_{n',n+1}|f(\psi',\psi)|^2\bigg]\,.
\end{split}
\end{equation}
Now that the two sectors are mixed, the overlap function is the sum of Eqs.~(\ref{eqn:OverlapA}) and~(\ref{eqn:OverlapB}),
\begin{equation}\label{eqn:Overlap3D}
f(\psi,\psi')=\sqrt{3}\alpha_1\alpha'_2-\alpha_2\alpha'_3+\alpha_4\alpha'_5-\sqrt{3}\alpha_5\alpha'_6\,.
\end{equation}

The 3D Kane magneto-optics can be broken down into nine contributions due to two three-fold distinctions.  The first includes intra- (AA and BB) and inter-sector transitions (AB), making up the set $\zeta=\{$AA, BB, AB$\}$.  The second distinction made pertains to the branches involved: flat($n$)--cone($n+1$) transitions ($\chi=1$), flat($n+1$)--cone($n$) transitions ($\chi=2$), and all cone--cone transitions ($\chi=3$), forming the set $\chi=\{$1, 2, 3$\}$.  Based on the symmetry of the wavefunctions (Eq.~(\ref{eqn:LLwavefunction})) and their overlap (Eq.~(\ref{eqn:Overlap3D})), inter-sector transitions ($\zeta=$AB) with the A state at Fock number $n$ will always have the B state at $n+1$, independent of which is the final or initial state in the transition.  A collection of snowshoe diagrams illustrating each of these nine sets of transitions is provided in Fig.~\ref{fig:SnowshoeParts}.  Each panel in the figure shows all transitions in the set, uninhibited by Pauli blocking.  For a given $\mu$, only those arrows that cross the energy $\eps=\mu$ will remain in the diagram.  To aid in the illustration, sector A is colored red, sector B in blue, and branches not involved in a particular set are shaded out to gray in each panel.  Note that the diagrams in Fig.~\ref{fig:SnowshoeParts} are at $k_z=0$ where LL's can be sorted into sectors A and B.  However, at this value of momentum, AB transitions cannot occur and the diagram is only used to illustrate the transitions that occur between levels integrated across $k_z$.

\begin{figure}
\begin{center}
\includegraphics[width=1.0\linewidth]{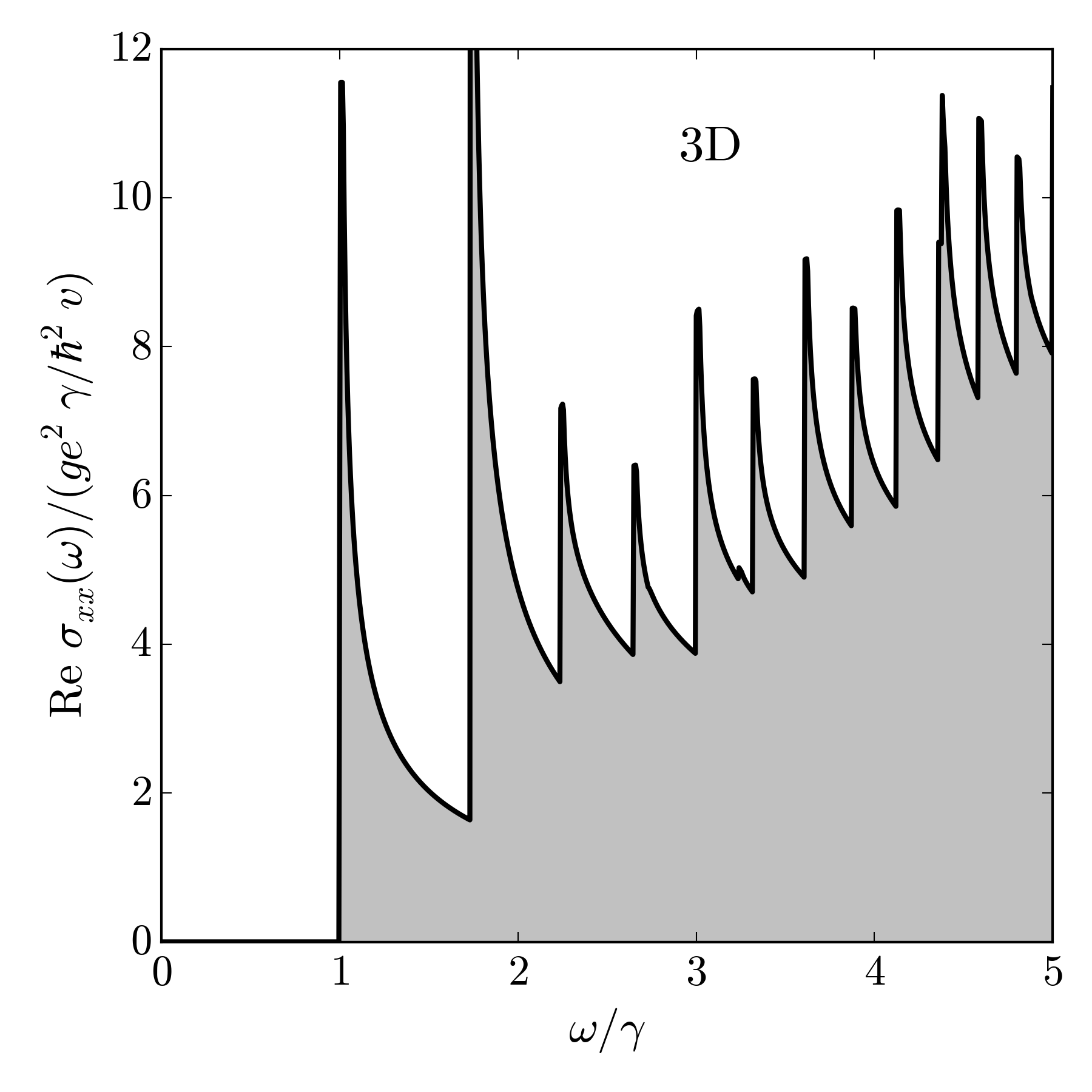}
\end{center}
\caption{\label{fig:Cond3D}Total longitudinal conductivity of the 3D Kane model with $T=0$, $E_g=0$, and $\mu=\gamma/2$ constructed using Eq.~(\ref{eqn:Cond3DSum}).}
\end{figure}

With the labelling system defined above, the conductivity in question is the summation
\begin{equation}\label{eqn:Cond3DSum}
\left.
\begin{aligned}
\Re \sigma^{\rm 3D}_{xx}(\om) \\
\Im \sigma^{\rm 3D}_{xy}(\om)
\end{aligned}
\right\}
=\frac{ge^2\gamma}{8\pi\hbar^2v}\sum_{\zeta}\sum_{\chi}\big(\zeta\chi\big)_{\substack{\\xx\\xy}}\,.
\end{equation}
The full results for the 3D Kane magneto-optical conductivity are
\begin{widetext}
\begin{equation}\label{eqn:AA1}
\begin{split}
& \big(AA1\big)_{\substack{\\xx\\xy}} = 3\sum_{\xi=\pm}\Nf(1)\Th(\bom+\xi\Eg/2)\Bigg(\frac{\Th(\bom^2+\xi\Eg\bom-1)}{(\bom^2+\xi\Eg\bom+2)\sqrt{\bom^2+\xi\Eg\bom-1}} \\
& \phantom{xxxxxxxxxxxxxxxxxxxx} +\frac{1}{\bom^2+\xi\Eg\bom-4}\sum_{n=0}^{\left\lfloor\frac{1}{4}(\bom^2+\xi\Eg\bom-9)\right\rfloor}\frac{(n+1)(n+2)}{[\bom^2+\xi\Eg\bom-(3n+7)]\sqrt{\bom^2+\xi\Eg\bom-(4n+9)}}\Bigg)\,,
\end{split}
\end{equation}
\begin{equation}
\big(AA2\big)_{\substack{\\xx\\xy}} = 3\sum_{\xi=\pm}\Nf(2)\frac{\Th(\bom-\xi\Eg/2)}{\bom^2-\xi\Eg\bom+4}\sum_{n=0}^{\left\lfloor\frac{1}{4}(\bom^2-\xi\Eg\bom-1)\right\rfloor}\frac{\bom^2-\xi\Eg\bom-(3n-1)}{\sqrt{\bom^2-\xi\Eg\bom-(4n+1)}}\,, \phantom{xxxxxxxxxxxxxxxxxxxxx}
\end{equation}
\begin{equation}
(AA3)_{\substack{\\xx\\xy}} = \frac{4}{\bom}\sum_{\xi=\pm}\Nf(3_{\rm tra})\frac{(\bom^2+2\xi\Eg\bom-8)^2}{\big|\bom^4-(4-\xi\Eg\bom)^2\big|}\sum_{n=0}^{\left\lfloor\frac{\bom^4-(\Eg^2+12)\bom^2+16}{16\bom^2}\right\rfloor}\frac{n+1}{\sqrt{\bom^4-4\big[(\Eg/2)^2+4n+3\big]\bom^2+16}}\,, \phantom{xxxxx}
\end{equation}
\begin{equation}
\begin{split}
& (BB1)_{\substack{\\xx\\xy}} = 3\sum_{\xi=\pm}\Nf(1)\Th(\bom+\xi\Eg/2)\Bigg(\frac{\Th(\bom^2+\xi\Eg\bom-3)}{\sqrt{\bom^2+\xi\Eg\bom-3}} \\
& \phantom{xxxxxxxxxxxxxxxxxxxxxxxxxxxxxxxxxxx} +\frac{1}{\bom^2+\xi\Eg\bom-4}\sum_{n=0}^{\left\lfloor\frac{1}{4}(\bom^2+\xi\Eg\bom-11)\right\rfloor}\frac{\bom^2+\xi\Eg\bom-(3n+10)}{\sqrt{\bom^2+\xi\Eg\bom-(4n+11)}}\Bigg)\,,
\end{split}
\end{equation}
\begin{equation}
(BB2)_{\substack{\\xx\\xy}} = 3\sum_{\xi=\pm}\Nf(2)\frac{\Th(\bom-\xi\Eg/2)}{\bom^2-\xi\Eg\bom+4}\sum_{n=0}^{\left\lfloor\frac{1}{4}(\bom^2-\Eg\bom-3)\right\rfloor}\frac{(n+1)(n+2)}{[\bom^2-\xi\Eg\bom-(3n+2)]\sqrt{\bom^2-\xi\Eg\bom-(4n+3)}}\,, \phantom{xxxx}
\end{equation}
\begin{equation}\label{eqn:BB3}
\begin{split}
& (BB3)_{\substack{\\xx\\xy}} = \frac{4}{\bom}\sum_{\xi=\pm}\Nf(3_{\rm tra})\frac{(\bom^2-2\xi\Eg\bom+8)^2}{|\bom^4-(4-\xi\Eg\bom)^2|}\sum_{n=0}^{\left\lfloor\frac{\bom^4-(\Eg^2+20)\bom^2+16}{16\bom^2}\right\rfloor}\frac{n+1}{\sqrt{\bom^4-4\big[(\Eg/2)^2+4n+5\big]\bom^2+16}}\,, \phantom{xxxxx}
\end{split}
\end{equation}
\begin{equation}\label{eqn:AB1}
(AB1)_{\substack{\\xx\\xy}} = 3\sum_{\xi=\pm}\Nf(1)\frac{\Th(\bom+\xi\Eg/2)}{\bom^2+\xi\Eg\bom+2}\sum_{n=0}^{\left\lfloor\frac{1}{4}(\bom^2+\xi\Eg\bom-5)\right\rfloor}\frac{(n+2)\sqrt{\bom^2+\xi\Eg\bom-(4n+5)}}{\bom^2+\xi\Eg\bom-(3n+4)}\,, \phantom{xxxxxxxxxxxxxxx}
\end{equation}
\begin{equation}
(AB2)_{\substack{\\xx\\xy}} = 3\sum_{\xi=\pm}\Nf(2)\frac{\Th(\bom-\xi\Eg/2)}{\bom^2-\xi\Eg\bom-2}\sum_{n=0}^{\left\lfloor\frac{1}{4}(\bom^2-\xi\Eg\bom-7)\right\rfloor}\frac{(n+1)\sqrt{\bom^2-\xi\Eg\bom-(4n+7)}}{\bom^2-\xi\Eg\bom-(3n+5)}\,, \phantom{xxxxxxxxxxxxxxx}
\end{equation}
\begin{equation}\label{eqn:AB3}
(AB3)_{\substack{\\xx\\xy}} = \bom\sum_{\xi=\pm}\Nf(3_{\rm ter})\sum_{n=0}^{\left\lfloor\frac{\bom^4-(\Eg^2+8)\bom^2+4}{16\bom^2}\right\rfloor}\frac{\sqrt{\bom^4-4\big[(\Eg/2)^2+4n+2\big]\bom^2+4}}{|\bom^4-(2+\xi\Eg\bom)^2|}\,. \phantom{xxxxxxxxxxxxxxxxxxxxxx}
\end{equation}
\end{widetext}
In Eqs.~(\ref{eqn:AA1})-(\ref{eqn:AB3}), $\Th(x)$ is the Heaviside step function and the joint-density factor
\begin{equation}\label{eqn:Nf}
\Nf(\chi) = \del_{\xi,+}\big[\nf(x_1)-\nf(x_2)\big]\pm\del_{\xi,-}\big[\nf(x_3)-\nf(x_4)\big]
\end{equation}
is different for each of the branch hopping types: 1, 2, and 3.  Further, this factor for the cone-cone transitions ($\chi=3$) is different for intra-system (AA,BB; $3_{\rm tra}$) and inter-system transitions (AB; $3_{\rm ter}$) as well.  The values of $x_1$--$x_4$ in Eq.~(\ref{eqn:Nf}) for each case are summarized in Table~\ref{tab:Nf}.  The $\pm$ in Eq.~(\ref{eqn:Nf}) corresponds to the longitudinal ($\Re\sigma_{xx}$) and transverse ($\Im\sigma_{xy}$) conductivity, respectively.  This is the only distinction between the two conductivities in Eqs.~(\ref{eqn:AA1})-(\ref{eqn:AB3}).

\begin{figure}
\begin{center}
\includegraphics[width=1.0\linewidth]{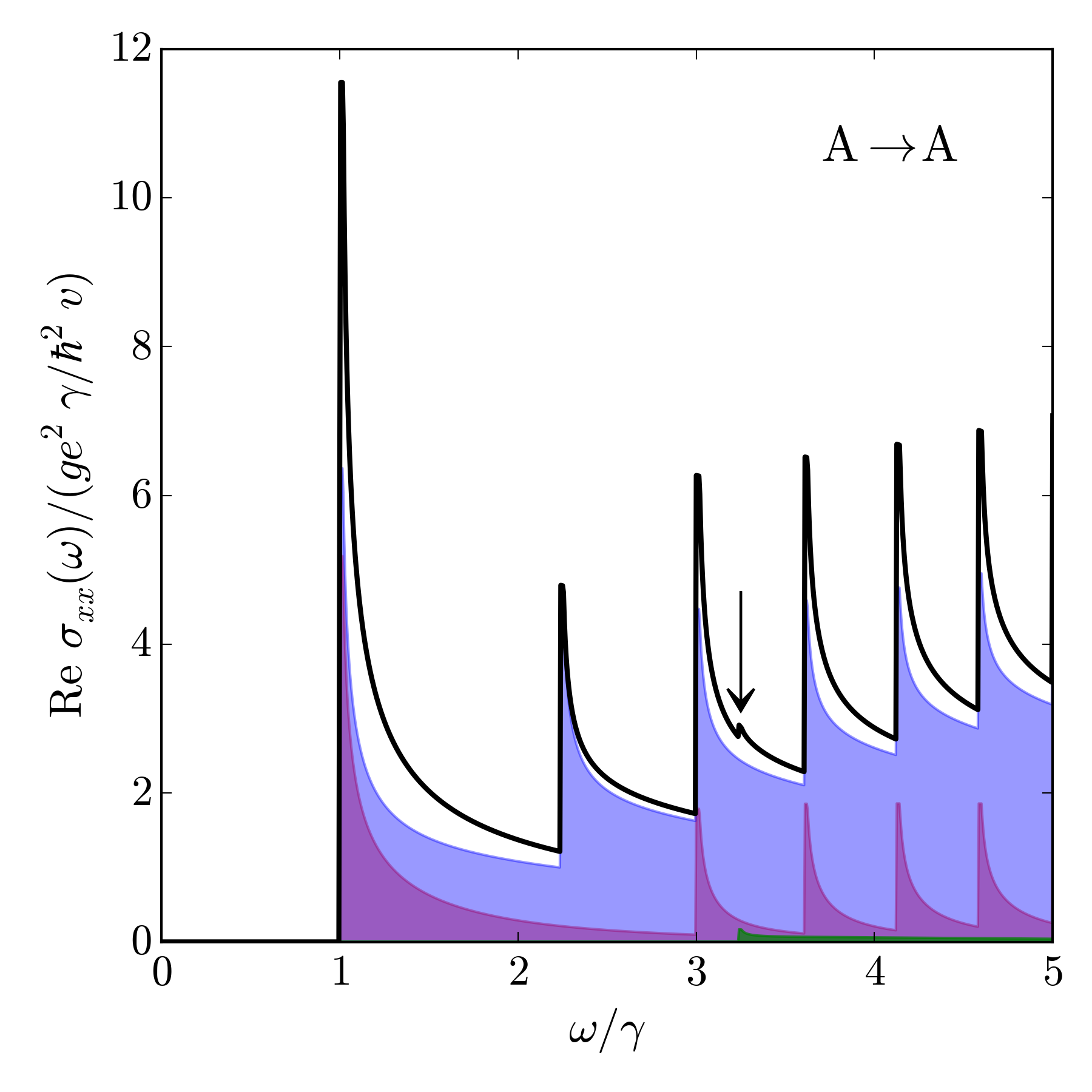}
\end{center}
\caption{\label{fig:Cond3DAA}(Color online) Contribution from AA-type transitions (black) to the total conductivity (Fig.~\ref{fig:Cond3D}).  This is further broken down into $\chi=1,2,3$-type transitions shown in red, blue, and green, respectively.  These colors are mixed when overlapping.  Indicated with an arrow is the location of the single $\chi=3$ peak on this energy scale.}
\end{figure}

\begin{figure}
\begin{center}
\includegraphics[width=1.0\linewidth]{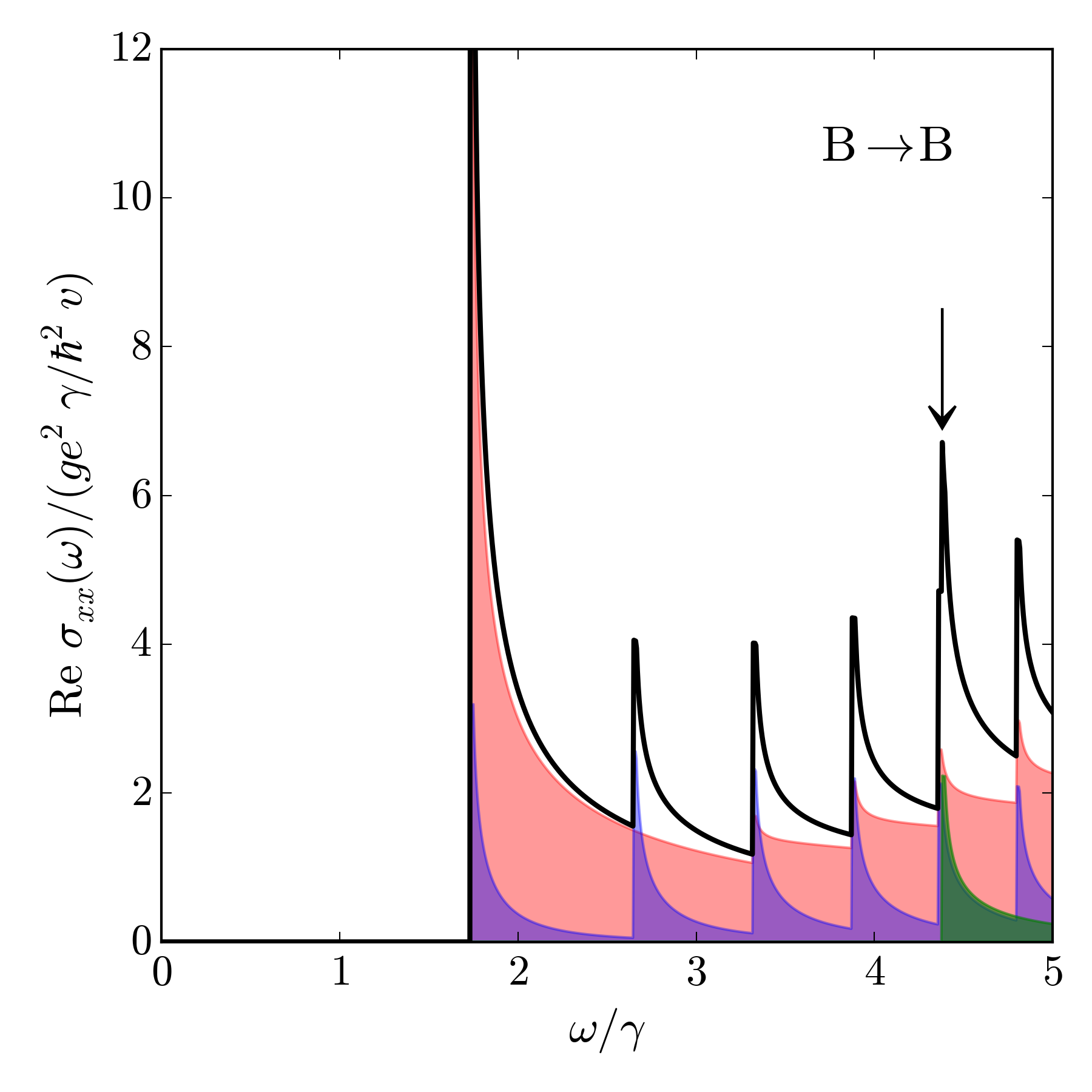}
\end{center}
\caption{\label{fig:Cond3DBB}(Color online) Contribution from BB-type transitions (black) to the total conductivity (Fig.~\ref{fig:Cond3D}).  The coloring of further dissections is as in Fig.~\ref{fig:Cond3DAA}.  Indicated with an arrow is the single $\chi=3$ on this energy scale, which closely overlaps another peak by chance.}
\end{figure}

\begin{figure}
\begin{center}
\includegraphics[width=1.0\linewidth]{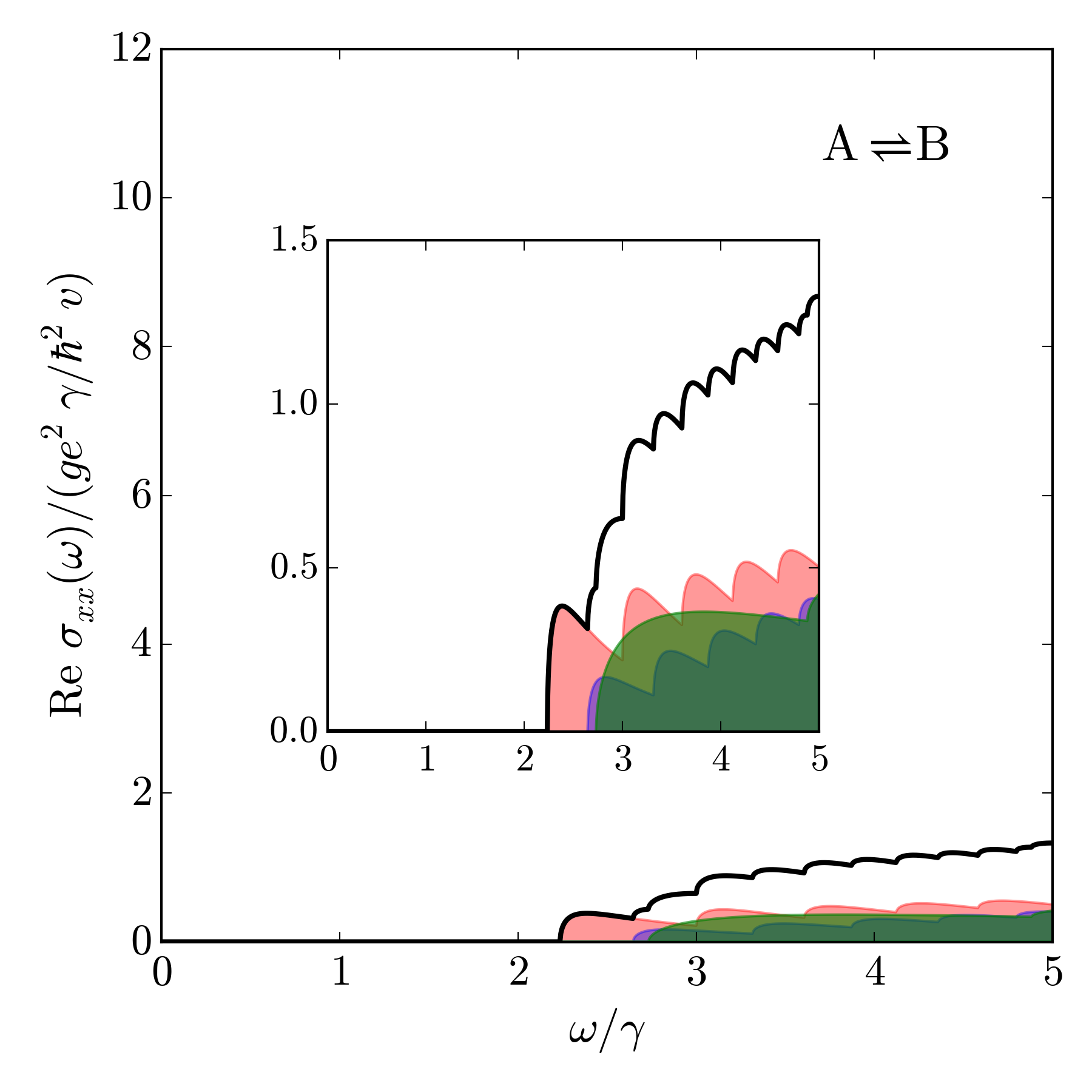}
\end{center}
\caption{\label{fig:Cond3DAB}(Color online) Contribution from AB-type transitions (black) to the total conductivity (Fig.~\ref{fig:Cond3D}).  The coloring of further dissections is as in Fig.~\ref{fig:Cond3DAA}.  Inset is the same but on a smaller vertical scale.}
\end{figure}

\begin{table}[t]
\caption{\label{tab:Nf} Values of $x_1$--$x_4$ to be used in $\Nf(\chi)$ (Eq.~(\ref{eqn:Nf})) for the different values of $\chi$.}
\begin{tabular}{c c c c c}
$\chi$ &\hspace{10pt} 1 &\hspace{10pt} 2 &\hspace{10pt} $3_{\rm tra}$ &\hspace{10pt} $3_{\rm ter}$ \\
\hline
\rule{0pt}{4ex} $x_1$ &\hspace{10pt} $-\bom$ &\hspace{10pt} 0 &\hspace{10pt} $\frac{\Eg}{2}-\frac{\bom^2+4}{2\bom}$ &\hspace{10pt} $\frac{\Eg}{2}-\frac{\bom^2-2}{2\bom}$ \\
\rule{0pt}{4ex} $x_2$ &\hspace{10pt} 0 &\hspace{10pt} $\bom$ &\hspace{10pt} $\frac{\Eg}{2}+\frac{\bom^2-4}{2\bom}$ &\hspace{10pt} $\frac{\Eg}{2}+\frac{\bom^2+2}{2\bom}$ \\
\rule{0pt}{4ex} $x_3$ &\hspace{10pt} 0 &\hspace{10pt} $-\bom$ &\hspace{10pt} $\frac{\Eg}{2}-\frac{\bom^2-4}{2\bom}$ &\hspace{10pt} $\frac{\Eg}{2}-\frac{\bom^2+2}{2\bom}$ \\
\rule{0pt}{4ex} $x_4$ &\hspace{10pt} $\bom$ &\hspace{10pt} 0 &\hspace{10pt} $\frac{\Eg}{2}+\frac{\bom^2+4}{2\bom}$ &\hspace{10pt} $\frac{\Eg}{2}+\frac{\bom^2-2}{2\bom}$ \\
\end{tabular}
\end{table}

The total longitudinal conductivity (Eq.~(\ref{eqn:Cond3DSum})) is plotted in Fig.~\ref{fig:Cond3D} for $T=0$, $E_g=0$, and $\mu=\gamma/2$, as in Fig.~\ref{fig:Cond2D} (the 2D case).  With this $\mu$, there are no intra-branch transitions present.  For a simpler presentation, the total conductivity is dissected into its three sector-type contributions: AA (Fig.~\ref{fig:Cond3DAA}), BB (Fig.~\ref{fig:Cond3DBB}), and AB (Fig.~\ref{fig:Cond3DAB}).  Within each of these latter three figures, the $\chi=1$-type contributions are in red, $\chi=2$ in blue, and $\chi=3$ in green.  These colors are in transparency so that overlapping regions mix (for example, blue and red mix to purple).  In Fig.~\ref{fig:Cond3DAB}, the contribution is shown inset on a smaller scale.  These inter-sector contributions occur on such a small scale (about ten times less than intra-sector) that they cannot be discerned from the total conductivity in Fig.~\ref{fig:Cond3D}.

The extrema of the dispersive LL's in 3D (Fig.~\ref{fig:SnowshoeProjection}) each occur at $k_z=0$.  For an individual transition between two LL's, as $k_z$ is traced out away from the origin, the separation in energy between two levels in any inter-branch transition increases.  It is the piece at $k_z=0$ which has the strongest contribution to the integral in Eq.~(\ref{eqn:Cond3DKane}), leading to the square-root singularities present in Eqs.~(\ref{eqn:AA1})-(\ref{eqn:BB3}) and illustrated in Figs.~\ref{fig:Cond3D}-\ref{fig:Cond3DBB}.  The $k_z=0$ piece lies at the lowest possible value of the energy separation $\om$, so that the singluarities occur at the exact location of the Lorentzian peaks in Fig.~\ref{fig:Cond2D} and then trail off toward higher energy.  The addition of the tails from successive peaks adds to an overall linear trend, which is to be expected for the conductivity of a relativistic system in 3D.\cite{Timusk13}  If there was an intracone transition, the associated peak would trail off to the left as tracing out $k_z$ from the origin decreases the separation in the energy levels involved.  Ashby and Carbotte provide a good analysis of the change in peak location with changing $\mu$ for another 3D relativistic system, the pseudospin-1/2 Weyl semimetal.\cite{Ashby13}  The patterns involved in the Kane system are much the same as in the Weyl system, both being relativistic in nature, and so are not discussed here.

\begin{figure}
\begin{center}
\includegraphics[width=1.0\linewidth]{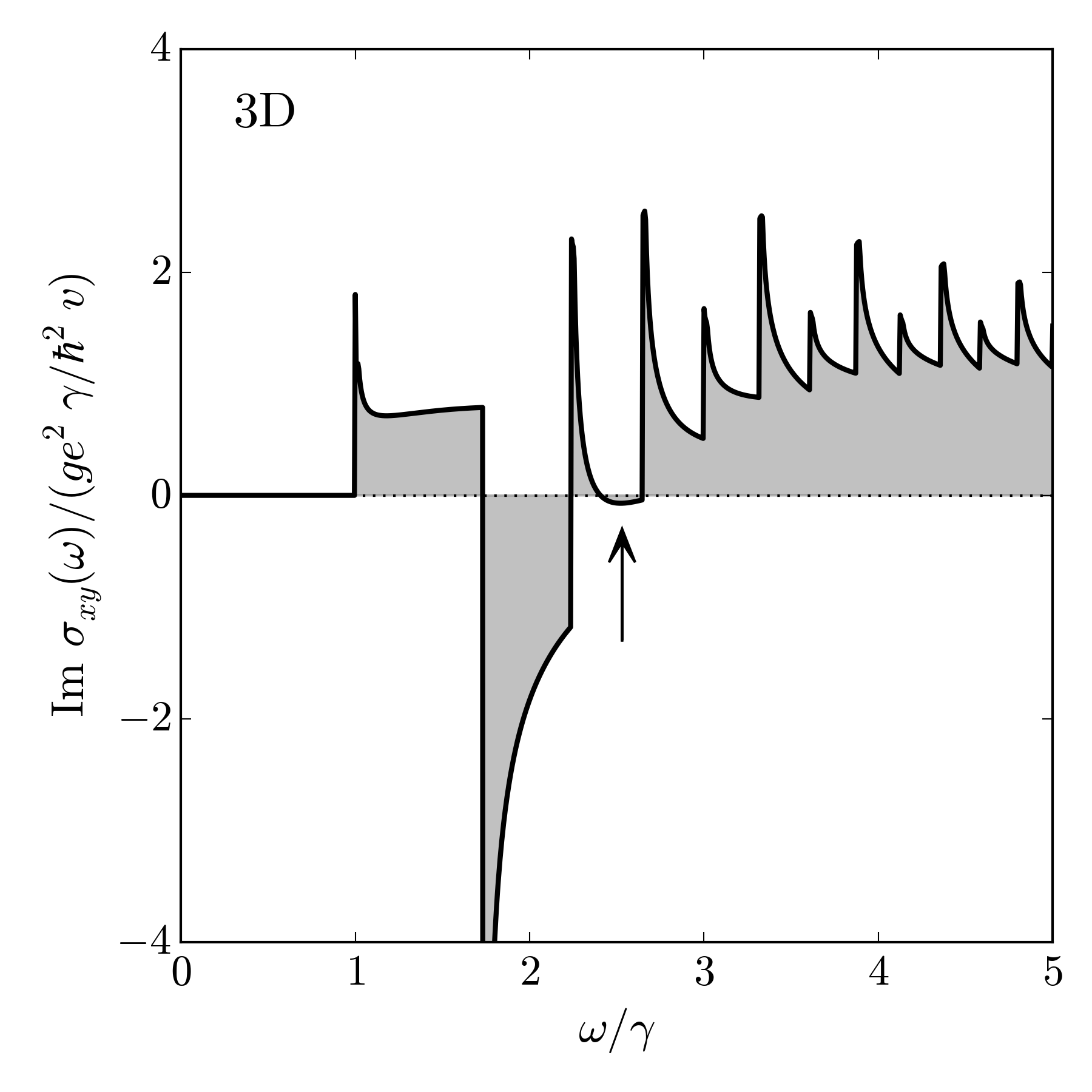}
\end{center}
\caption{\label{fig:CondXY} Total transverse conductivity of the 3D Kane model with $T=0$, $E_g=0$, and $\mu=\gamma/2$.  Indicated is a dip to negative values which is a signature of inter-sector transitions occuring.}
\end{figure}

At the energy scale shown, only one intercone (green) peak is present in each of Figs.~\ref{fig:Cond3DAA}-\ref{fig:Cond3DAB}.  As in the 2D Kane conductivity in Sec.~\ref{sec:2D} and for the three-band relativistic system in general, the flat-cone contributions are larger than the intercone contributions for the intra-sector transitions (AA and BB).  In general, peaks are larger for transitions between levels with smaller separation in energy and for transitions occuring at a lower Fock number.  The intercone peaks in Figs.~\ref{fig:Cond3DAA} and~\ref{fig:Cond3DBB} are indicated with an arrow.  While this peak in the latter figure appears a higher energy, it is still larger than that in the former because the peak has a smaller value of $n$ associated with it.

\begin{figure}
\begin{center}
\includegraphics[width=1.0\linewidth]{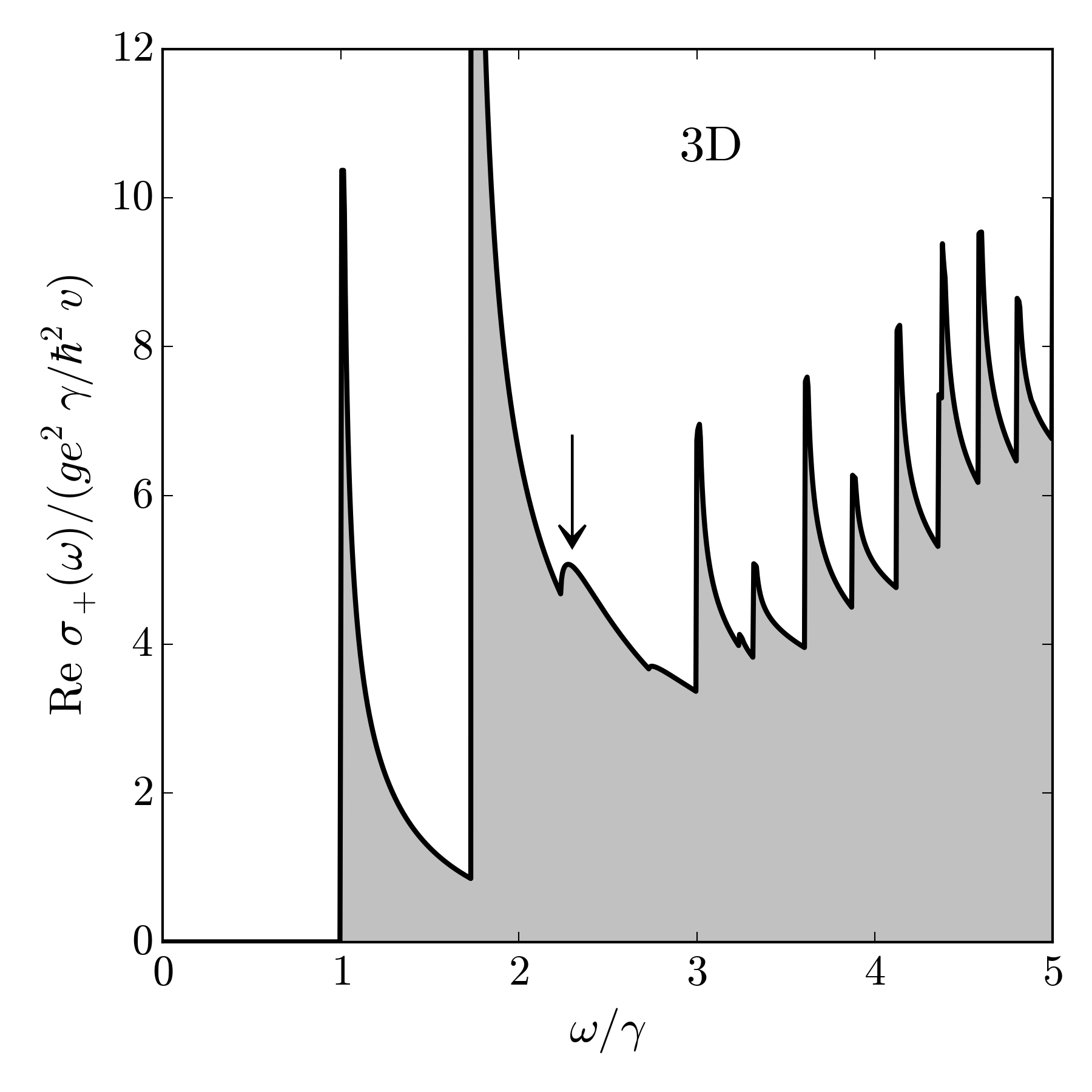}
\end{center}
\caption{\label{fig:CondPos} Optical conductivity of the 3D Kane model for light with right-handed circular polarization with $T=0$, $E_g=0$, and $\mu=\gamma/2$.  Indicated is a peak arising only from inter-sector transitions.}
\end{figure}

The discussion of the square-root singularities that are found in Figs.~\ref{fig:Cond3D}-\ref{fig:Cond3DBB} only pertains to the AA and BB transitions.  Because AB transitions cannot occur at $k_z=0$, the singular behaviour is suppressed.  Instead these contributions have a regular square-root profile seen in Eqs.~(\ref{eqn:AB1})-(\ref{eqn:AB3}) and Fig.~\ref{fig:Cond3DAB}, but which are too minor to be seen in the full conductivity.  Square-root profiles like this are seen in the theoretical modelling of a magneto-optical study on ZrTe$_5$, which has been identified as a 3D Dirac semimetal.\cite{Chen15a,Chen15b}

Where the effects of the inter-sector contributions cannot be discerned from the longitudinal conductivity, there is an impact on the transverse conductivity ($\Im\sigma_{xy}$), plotted in Fig.~\ref{fig:CondXY}.  In the region of $\omega\sim2.5\gamma$, the curve dips below zero, indicated with an arrow.  If the inter-sector contribution is not included, this dip does not reach negative values.  This gives a clear indication of communication between the sectors in the 3D conductivity which does not occur in 2D.  Also in this plot, any cone-cone contribution is exactly zero.  With the value $E_g=0$, the low-energy Kane model exhibits particle-hole symmetry, which is reflected in the overlap function by the identity
\begin{equation}
\big|f(\psi^{\zeta}_{\xi,n+1};\psi^{\zeta'}_{\xi',n})\big|^2 = \big|f(\psi^{\zeta}_{-\xi,n+1};\psi^{\zeta'}_{-\xi',n})\big|^2\,.
\end{equation}
Visually, this means that all transitions in the snowshoe diagram that are mirror symmetric about $\epsilon=0$ have the same squared-amplitude overlap function.  The transitions in each pair have opposite polarization: where one is from a level $n\rightarrow n+1$, its conjugate will be from $n+1\rightarrow n$.  The negative sign found in Eq.~(\ref{eqn:Nf}) for the transverse conductivity means then that these mirror-symmetric pairs cancel each other out in the transverse conductivity due to the opposite polarizations.  This is exactly the case of all cone-cone transitions for $\mu=\gamma/2$ in Fig.~\ref{fig:CondXY} (see Fig.~\ref{fig:Snowshoe2D} which shows mirror-symmetric cone-cone transitions in the 2D snowshoe diagrams).

\begin{figure}
\begin{center}
\includegraphics[width=1.0\linewidth]{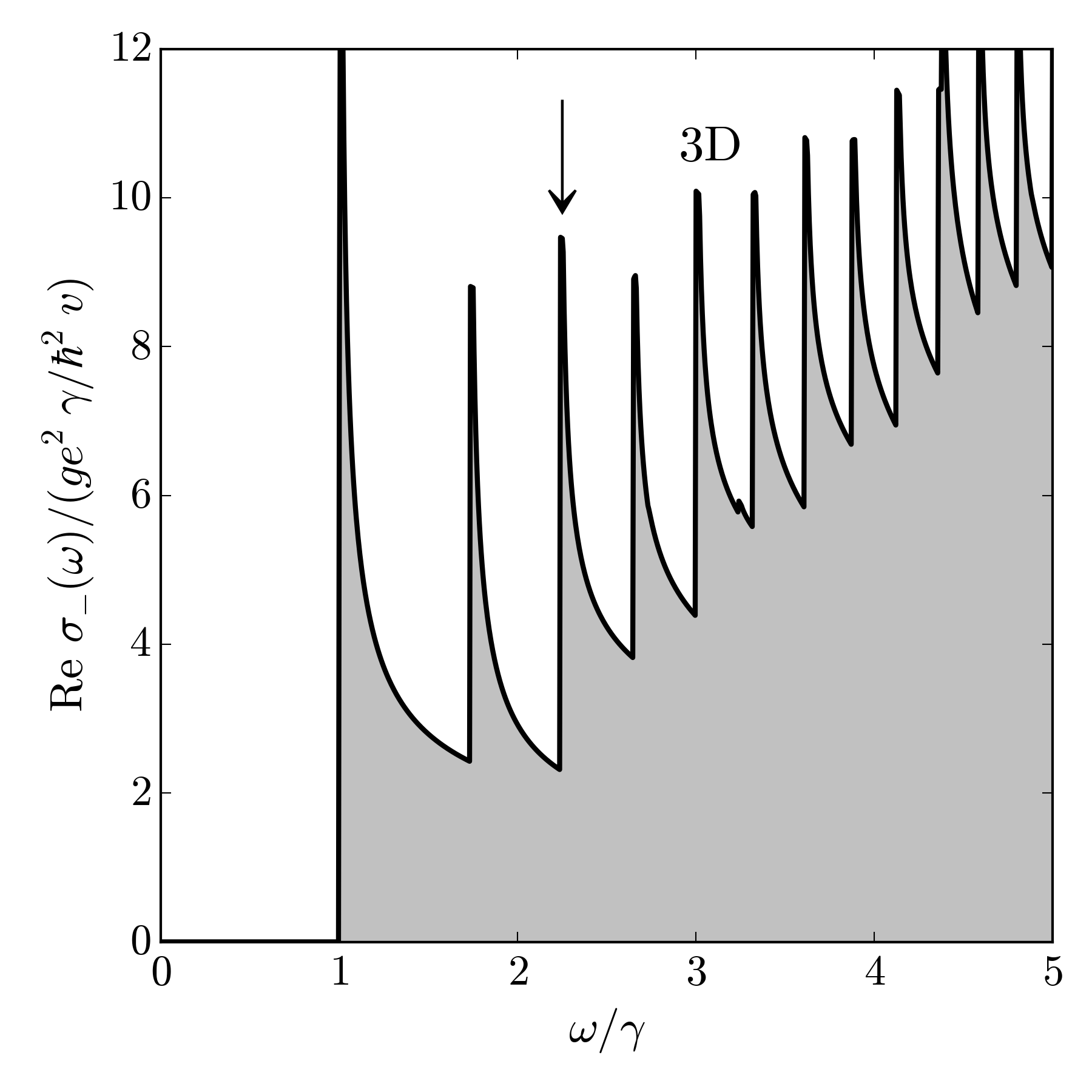}
\end{center}
\caption{\label{fig:CondMin} Optical conductivity of the 3D Kane model for light with left-handed circular polarization with $T=0$, $E_g=0$, and $\mu=\gamma/2$.  Indicated is a peak arising only from intra-sector transitions.}
\end{figure}

The inter-sector contributions can also be pieced out from the conductivity for circularly polarized light.  For right- and left-handed circular polarizations,
\begin{equation}\label{eqn:Circular}
\Re\sigma_{\pm} = \Re\sigma_{xx}\mp\Im\sigma_{xy}.
\end{equation}
These are plotted in Figs.~\ref{fig:CondPos} and~\ref{fig:CondMin}, respectively.  Only those transitions with $n\rightarrow n+1$ contribute to $\Re\sigma_+$ and only $n+1\rightarrow n$ to $\Re\sigma_-$.  Because of the lack of a Landau level in the flat branch at $n=2$ in sector A and $n=1$ in sector B, there are two positive-energy Landau levels in which transitions from the flat band are polarized so that intra-sector transitions are directed $n+1\rightarrow n$, and the inter-sector transitions are $n\rightarrow n+1$.  Then, by observing the circular polarized conductivities, these special contributions can be separated out.  In Fig.~\ref{fig:CondPos}, there are two such peaks where only the inter-sector contribution is observed, the first of which is indicated with an arrow.  The intra-sector singular contribution is relegated to Fig.~\ref{fig:CondMin}, where it cannot drown out the non-singular inter-sector piece.  Experimental measurement of the conductivity for circular polarizations could reveal inter-sector transitions and could also be used to investigate the transverse conductivity (Fig.~\ref{fig:CondXY}) by rearranging Eq.~(\ref{eqn:Circular}) in the proper way.

\section{Conclusions}

As detailed in Sec.~\ref{sec:Intro}, the Kane model for zinc-blende semiconductors is coming to the front line of relativistic materials research due to observations of massless Kane fermions.  To supplement the discussions in this field, we have provided analytic expressions for the magneto-optical conductivity of the low-energy Kane model in both 2D and 3D (Eqs.~(\ref{eqn:Cond2DSum}) and~(\ref{eqn:Cond3DSum}), respectively).  Results are presented in Figs.~\ref{fig:Cond2D} and~\ref{fig:Cond3D} for the longitudinal component of the conductivity ($\Re\sigma_{xx}$) and for specific values of $T=0$, $E_g=0$, and $\mu=\gamma/2$.  Other components---the transverse conductivity ($\Im\sigma_{xy}$) and for circular polarizations ($\Re\sigma_+$ and $\Re\sigma_-$)---are presented in Figs.~\ref{fig:CondXY}, \ref{fig:CondPos}, and \ref{fig:CondMin}, respectively.  We see that in these latter presentations, there is evidence of the inter-sector transitions not present in 2D chiral, or relativistic, condensed matter systems.  While we chose to present conductivities for the simplest values of $T$, $E_g$, and $\mu$, each parameter can be varied, allowing for versatile application of our analytical results.

This research has been supported by the Natural Sciences and Engineering Research Council of Canada.  We would like to thank E. Illes for helpful discussions.

\bibliography{bibliography}
\end{document}